# Object kinetic Monte Carlo simulations on the difference between fission neutron and heavy ion irradiation induced void evolution in Fe-Cr alloys

Bowen Zhang[1], Fengping Luo[1], Yuxin Liu[1], Jin Wang[1], Denghuang Chen[1], Xun Guo[2], Chenxu Wang[1,*], Steven J. Zinkle[3], Yugang Wang[1,*]

[1]*State Key Laboratory of Nuclear Physics and Technology, Center for Applied Physics and Technology, Peking University, Beijing 100871, China*
[2]*Advanced Research Institute of Multidisciplinary Science, Beijing Institute of Technology, Beijing 100081, China*
[3]*Department of Nuclear Engineering, University of Tennessee, Knoxville, TN, 37996, USA*

Abstract: Experimental results show significant difference between neutrons and ions irradiation of alloys, while the underlying reasons remain unclear. Herein, we performed object kinetic Monte Carlo (OKMC) simulations on void evolution in Fe-Cr alloys under neutron and ion irradiations, focusing on the effects of dose rate, irradiation particle type and temperature. Binary Collision Approximation and Molecular Dynamics are applied to obtain the cascade morphology of ion irradiation in order to study the effect of spatial correlation of cascades along the ion track, which is considered as a significant difference between the neutron and ion irradiations. Systematic OKMC simulations were performed at a wide range of dose rate from $10^{-7}$ to $10^{-3}$ dpa/s and temperature from 300 to 500°C. Simulation results show that both a higher dose rate and a lower temperature can lead to a higher density and a smaller average size of voids. High dose rate greatly promotes the interaction frequency between small defects and inhibit the absorption of vacancies by one vacancy cluster, thus enhancing the nucleation of vacancy clusters. This dose rate effect explains the major difference of microstructure between fission neutron and heavy ion irradiation. High temperature enhances the migration of small defects and the absorption of vacancies by vacancy clusters, and thus enlarge the vacancy clusters. The impact of irradiation particle types that has influence on the primary knock-on atom spectrum and cascade morphology is less important to void evolution compared with dose rate and irradiation temperature. This work provides fundamental insights into the difference in void evolution between fission neutron and heavy ion irradiations that are important to the application of ion irradiations in study irradiation effects of nuclear materials.

**Keywords**: OKMC, neutron and ion irradiation, dose rate, cascade morphology, temperature

## 1. Introduction

The development of advanced nuclear energy systems presents a promising solution to the global energy crisis and climate change. These reactors operate under extremely harsh conditions, characterized by high temperatures and intense radiation fields, necessitating structural materials with greater radiation tolerance than those used in current fission reactors to resist radiation-induced degradation [1]. However, evaluating the radiation effects on the microstructure and mechanical properties of these materials is both time-consuming and resource-intensive, often requiring years of research and hampered by the limited availability of test reactors. Therefore, for decades, efforts have been made to search for methods to achieve high damage level in a short time, enabling predictions of microstructure and mechanical property changes at a lower cost [2-5].

The study of neutron radiation damage using heavy ion irradiation has emerged as a promising method due to the convenience of achieving high flux irradiation with effective temperature control [6-8]. Ion irradiation is thus widely employed for comparative studies to investigate the effect of irradiation conditions, such as damage rate and temperature [9]. Previous studies have shown that heavy ion irradiation can produce microstructure (i.e. voids, dislocation loops and precipitation) similar to those under neutron irradiation in Fe-based alloys. However, a key difference between the two irradiation conditions is the density and size of the defects, leading to a difference in macroscopic void swelling and hardening [10]. Quantitative analysis of the microstructure data between neutron and ion irradiation conditions requires considering some main difference of neutron and ion irradiation: the dose rate, primary knock-on atom (PKA) spectrum, cascade morphology induced by neutron and ion atoms, the transmutation gas He, and the limited depth of penetration of ion irradiation [11-14]. There have been numerous studies to describe and clarify the different irradiation effects between neutron and ion irradiation. In the early 1970s, Mansur et al. attributed the difference to the dose rate and developed a model that changes of dose rate can be accommodated by a shift in temperature to keep the overall defect aggregation properties (such as swelling) invariant [15, 16]. Was *et al.* applied this model successfully in case of irradiation effects induced by protons and neutrons at low doses (less than 5 dpa) and low dose rates (~$10^{-6}$ dpa/s) [9, 17]. More recently, object kinetic Monte Carlo simulations were also performed to investigate the influence of dose rates, and suggested that lower dose rate results in an increase in the dislocation loop density and thus enhances irradiation hardening [18, 19]. The effects of transmutation gases and impurity elements have also been extensively investigated over the long-term investigation. Was et al. [20, 21] focused on the impact of transmutation gas He and H which are significantly higher in fusion reactor than fission reactor or ion irradiation. Their findings highlight the complicated synergies between He, H and radiation damage of structural materials [22]. Shao et al. [23, 24] observed that the impurities such as carbon atoms introduced by accelerator-based ion irradiation play a significant role in the void swelling.

Structural materials experience severe property degradation under neutron irradiation, including radiation induced volume swelling, hardening, embrittlement, creep, phase instability [25]. Void swelling is a critical phenomenon of material degradation and is often considered as one of the most important criteria for the evaluation of materials radiation resistance property [26]. Moreover, void swelling is easily observed and measured in the case of the short range of ion irradiation. As a result, it is a meaningful method to compare the effect of neutron and ion irradiation by the swelling rates.

Ferritic/martensitic steels are considered promising candidates structure materials of the breeding blanket, reactor pressure vessel (RPV), as well as the fuel cladding in the future fusion and fission reactors [27]. High-chromium F/M steels are often employed as model materials in irradiation experiments and subsequent characterization for superior thermal and mechanical properties compared to austenitic steels, such as higher resistance to radiation-induced swelling [28, 29]. Consequently, Fe-Cr alloys serve as important model materials for mesoscopic simulations to enhance the knowledge of the microstructure evolution and macroscopic mechanical properties of structural materials under neutron and ion irradiations [30, 31].

Sustained efforts have been made to investigate the correlation between primary damage and long-term microstructural evolution using various methods including molecular dynamics and kinetic Monte Carlo. Zinkle et al. [32] found that at temperatures where interstitials are mobile, the surviving defect fraction of displacements is similar over a wide range of PKA energies for various ion and fast neutron irradiation conditions considering correlated recombination. M. Hou et al. [33] using object kinetic Monte Carlo simulations, demonstrated that cluster size distributions in cascade debris and the spatial extent of vacancy and SIA clusters in displacement cascades play major role in the evolution of cluster size distributions after long enough time (at 0.1 dpa). In contrast, other properties of displacement cascades, such as sub-cascade formation or radial pair correlation functions play relatively minor role.

Moreover, one significant difference between the neutron and ion irradiation rarely considered before is the spatial distribution of the PKA and cascades. In most cases of the previous OKMC simulations, the cascades are spatially randomly introduced in the whole simulation box [34-36]. However, the range of neutron is much longer than that of ion, and the displacements of neutron irradiation are more separated due to the longer mean free paths, compared to the heavy ion irradiation, which has more concentrated displacements along its track [11, 37]. This can make a difference in the spatial correlation between displacement cascades when they interact with each other. in this work, object kinetic Monte Carlo (OKMC) simulations were performed on Fe-9%Cr alloys to investigate the difference between neutron and Fe ion irradiations since OKMC could provide spatial information of primary defects. Both the difference in PKA energy spectrum and the spatial correlation of the primary knock-on atoms and displacement cascades of the neutron and ion irradiations were considered. In our simulation, the cascades of neutron were regarded as random distribution in the simulation box, while the spatial distribution of ion irradiation PKA and cascades is taken into consideration. The cascades of ion were introduced in the form of the total ion track caused by one incident ion to study the effect of correlation between the nearby cascades of ion track. The influences on the defect cluster microstructure evolution are investigated upon several different irradiation conditions including varied irradiation temperatures and dose rates. This work provides new insights into the differences between neutron and ion irradiations on irradiation microstructure evolution in F/M steels.

## 2. Methods

We used the open source software, MMonCa[38] for the object kinetic Monte Carlo simulation. The OKMC model contains only defects, such as vacancies/SIAs, their clusters and dislocation loops, while Fe atoms on the lattice site are neglected.

The core algorithm in the OKMC method is from the stochastic Monte Carlo algorithm described in [39, 40]. In the OKMC method, possible events for the objects occur randomly at each step to describe the evolution of the system. The events contain defect migration, cluster emission and de-trapping, and the rates of events can be calculated using Arrhenius relation, given as

$$\Gamma_{n,i} = v_{n.i} \exp(-\frac{E_{a,(n,i)}}{k_B T})$$

where $v_{n,i}$ and $E_{a,(n,i)}$ denotes the attempt frequency and activation energy for each object n and its related event i, respectively. $k_B$ and T are the Boltzmann constant, and temperature in Kelvin. In radiation damage simulations, damage production rate is considered as external events with the rate of $\Gamma^{ext}$. At each step of the OKMC simulation, an event is selected among all possible events based on their relative probabilities, which is proportional to the rates of events. Then the simulation time increment of each step follows the residence time algorithm, given as

$$\Delta\tau = -\ln(r)/(\sum_{n,i}\Gamma_{n,i} + \Gamma^{ext})$$

where r is a random number in the range of (0,1].

### 2.1 The parameters of the simulation box

The simulation box in our simulation is set to be $114.8\ \mathrm{nm} \times 60.27\ \mathrm{nm} \times 54.53\ \mathrm{nm}$ $(400a_0 \times 210a_0 \times 190a_0)$ with $a_0 = 0.287$ nm, where non-cubic shape is chosen in order to avoid potential anomalies from one-dimensional fast migration of <111> dislocation loops entering a migration trajectory loop, as discussed in [41]. And the longer x-direction of box ensures that the simulation box contains more displacement cascades caused by the ion track of one incident Fe ion. Similar to previous simulation for Fe-Cr alloy [31], the dislocation line density is $3.04 \times 10^{14}$ m$^{-2}$, and the effect of dislocation density is also studied below. Periodic boundary conditions (PBC) are applied in all three directions of the simulation box, and the grain boundaries are not considered in the current simulation [42]. The capture radius for point defects is set to be 0.287 nm [43] , and the capture volumes of clusters are the union of the capture distances of their constituent defects, depending on the shape of clusters, which is the basic settings in the MMonCa code. Bias absorption at the dislocation line is considered, and the bias factor for SIAs at dislocation lines is set as 1.05, for the capture efficiencies of SIAs are larger than vacancies [35].

### 2.2 The energetics of objects

OKMC method requires formation energy and migration energy of the objects contained in the simulation to be the inputs, in order to simulate dissociation, migration and emission reactions.

The migration energy barriers are specified in Table 1. In this work, single and small vacancy clusters migrate in 3D mode and vacancy clusters containing more than 5 vacancies are considered immobile, which is a common assumption in OKMC and cluster dynamics simulations [34]. Small interstitial clusters no more than 5 SIAs migrate in 3D and transform into 1/2<111> interstitial loops when it contains more than 5 SIAs. Two types of interstitial loops (1/2<111> and <100> type) experimentally observed are considered in the model. The 1/2<111> loops perform 1D migration along <111> directions and the <100> loops are immobile. The production of <100> loops are from the reaction between two 1/2<111> loops with comparable sizes, and further details can be seen in [44, 45].

**Table 1**

Migration energies ($E_m$) of the objects considered in the OKMC model. All values are from Ref. [46]

| Object groups | Sub-groups | $E_m$ (eV) |
|---|---|---|
| V Clusters | V | 0.67 |
| | V2 | 0.62 |
| | V3 | 0.35 |
| | V4 | 0.48 |
| | $V_n$, n>4 | immobile |
| I Clusters | I | 0.34 |
| | I2 | 0.42 |
| | I3 | 0.43 |
| | I4 | 0.43 |
| | $I_n$, n>4 | $0.06+0.11/n^{1.6}$ |
| 1/2<111> | | 0.1 |
| <100> | | immobile |

The formation energies of the objects are also essential parameters of the model. The MMonCa code requires formation energy as input parameters to calculate binding energy whenever dissociation or emission event occurs during the simulation, and the activation energy $E_{a,(n,i)}$ of these events equals to the binding energy plus the migration energy of the dissociated point defect. The formation energy of vacancy clusters is from a scaling model of our previous work to quantitatively evaluate the interaction energy of vacancy clusters in α-Fe and Fe-Cr ferritic alloys which is parameterized by ab-initio calculations [47]. The formation energy of interstitial clusters also comes from the ab-initio scaling laws referred to Ref [48]. The key parameters are specified in Table 2.

**Table 2**

Formation energies ($E_f$) of the objects considered in the OKMC model. All values are from Ref. [47, 48]

| Object groups | Sub-groups | $E_f$ (eV) |
|---|---|---|
| V Clusters | V | 2.07 |
| | $V_n$, n≥2 | $3.576\times n^{2/3} -1.617$ |
| I Clusters | I | 3.77 |
| | $I_n$, n≥2 | $3.77+(3.77-0.80)\times(n^{2/3}-1)/(2^{2/3}-1)$ |
| 1/2 <111> | $I_n$, n>4 | $1.60\sqrt{n}\ln n + 5.35\sqrt{n} - 0.15$ |
| <100> | $I_n$, n>4 | $1.78\sqrt{n}\ln n + 7.16\sqrt{n}$ - 5.82 |

In our simulation of the Fe-Cr alloys (Fe-9%Cr) , the effect of solute atoms, Cr, are not treated as objects for its small proportion. Instead, we consider its effect to reduce the mobility of SIA clusters, using a simplified approximation named “grey alloy” approach, as described in details in Ref. [31]. This approximation can avoid the introduction of an unnecessarily large number of objects of impurity atoms in the simulation box and accelerate the simulation.

### 2.3 The displacement cascades induced by neutron and ion irradiation

To clarify the difference of cascade correlation between neutron and heavy ion irradiations, we used the Binary Collision Approximation (BCA) method and the Molecular dynamics (MD) results to construct the cascade morphology of different irradiation particles. A connection between these two methods was first applied several decades ago by Jaraiz et al. [49] and was used to simulate high-energy collision cascades in solids under irradiation by Ortiz [50]. When coupled with OKMC models, this method could offer great efficiency with a considerable level of calculation accuracy [51].

In our simulation, the cascades of neutron were regarded as random distribution in the simulation box, while the cascades of ion were introduced in the form of the ion track caused by one incident ion considering the spatial distribution. However, for the Fe self-ion irradiation at MeV or higher energy levels, the relevant length scale is about several hundred or thousand nanometers, which goes beyond the practical reach of MD simulations [52]. In terms of this constraint, BCA method is used for the 5 MeV Fe ion irradiation. BCA simulations were performed by the Stopping and Range of Ions in Matter (SRIM) code [53] with the K-P model to obtain the spatial distribution of the primary knock-on atoms with varied energies of the 5 MeV ion irradiation. And the ion irradiation data of depth range 800-911.5 nm was chosen, which equals the length of the simulation box and is also the proper analysis region of the ion irradiation experiments (see Fig. 1).

The PKA spectrum of neutron irradiation was generated from typical neutron spectrum of HFIR reactor using the SPECTRA-PKA code [54], while the one for 5 MeV Fe ion irradiation was calculated by the SRIM code mentioned above at the region of 800-911.5 nm. Molecular dynamics simulations were performed to obtain a database of cascades with a discrete set of primary knock-on atom (PKA) energy values ranging from 0.1 keV to 200 keV. These PKA energies are used as a discrete approximation for a real, continuous spectrum found in neutron reactor or ion irradiation. Fig. 2. shows the approximate PKA energy spectra of the HFIR reactor and 5 MeV Fe ion irradiation, used as the neutron and ion irradiation spectra in our OKMC simulations. MD simulations were performed using the LAMMPS code [55]. The interatomic potential developed by Marinica et. al. [56, 57] (denoted M07) was used to describe the interaction between Fe atoms. Cascades were initiated by imparting a kinetic energy to a selected PKA energy in Fig.2 and an initial direction $\langle 135 \rangle$ was used for the PKA to avoid the channeling effect. The evolution of cascades was followed until a maximum time of 30 ps to obtain the cascade defects after the athermal recombination. Then the MD simulation cascades were used for certain atoms with discrete energies in Fig. 2. to reconstruct the ion track with spatial distribution at the range of 800-911.5 nm and to randomly insert the cascades as the neutron irradiation. The displacement cascades morphologies of different irradiation particles are shown in Fig. 2. Another type of simulation for ion irradiation with randomly distributed cascades (referred to as ion spectrum below) is also used as a reference, which only consider the impact of PKA spectra and ignore the spatial information. It is important to note that the three simulation types of irradiation conditions (neutron spectrum, ion track and ion spectrum) refer solely to their PKA spectra and spatial distribution of cascades, without representing their dose rate in our simulations. Moreover, OKMC simulations with dose rate ranging from $10^{-7}$ to $10^{-3}$ dpa/s were performed for all the three irradiation conditions, respectively.

The cascade defects are added to the simulation box by the irradiation dose rate. The dose of induced cascades was calculated by NRT-model [58], and the number of displacements ($N_d$) of each cascade can be obtained by

$$N_d = 0.8\, E_{PKA}/2E_d$$

where $E_d$ = 40 eV is the threshold displacement energy in Fe. The number of displaced atoms induced by cascades divided by the total number of the bcc lattice sites contained in the simulation box ($N_{atom}$) can be considered as the dose. And the dose when multiple cascades are introduced in the simulation box can be calculated by

$$\text{Dose} = 0.8 \sum E_{PKA} / (2E_d\, N_{atom})$$

## 3. Results

In this section, we analyze the results of irradiation microstructure evolution of our OKMC simulation under different irradiation conditions (dose rate, temperature and neutron/ion irradiation). All the simulation results were reported at the dose upto 1.0 dpa for dose rates $10^{-7}$ – $10^{-3}$ dpa/s in the simulation box. The simulations were performed at a large range of temperatures (300 °C, 400 °C and 500 °C, respectively). As mentioned above, the simulation box was set to be $400a_0 \times 210a_0 \times 190a_0$, $114.8\ \text{nm} \times 60.27\ \text{nm} \times 54.53\ \text{nm}$ ($a_0 = 0.287$ nm), and the minimal defect density (when only one defect is contained in the box) is thus $2.65 \times 10^{21}$ m$^{-3}$. Our goal in this work is to compared the neutron and ion irradiation by the evolution of voids. Their density is expectedly one order of magnitude higher than the SIA loops density, so a box with relatively smaller size can be chosen for higher computing efficiency. Every simulation was executed in 3 independent runs, allowing to perform statistics and thus reducing the minimal distinguishable density to $8.83 \times 10^{20}$ $m^{-3}$. This is still too high for performing proper simulations at 500 °C with $10^{-7}$ dpa/s in our model, so there is lack of statistical data at this irradiation condition. We mainly analyze the density, size and swelling of vacancy clusters for they are the major results of experimental observations.

### 3.1 Effect of dose rate

The number density and average diameter size of vacancy clusters with three types of irradiation conditions (neutron and ion spectrum with randomly distributed cascades; ion track with spatial distribution of cascades as mentioned in Section 2.3) under different irradiation temperatures and dose rates at 1.0 dpa are presented in Fig. 3, respectively. The results versus irradiation dose are shown in Fig. 4 and Fig. 5 with three different temperatures.

The impact of dose rate is similar among the three different types of irradiation conditions. As shown in Fig. 3, for the same type of irradiation condition and temperature, the density of vacancy clusters increases by one or two orders of magnitude with increasing dose rate from $10^{-7}$ to $10^{-3}$ dpa/s, while the average size of vacancy clusters decrease by 70% at most. The reason lies in the fact that at a higher dose rate, a larger number density of defects are introduced per unit time and unit volume, which enhances the interaction between irradiation defects [42]. This leads to a larger density of nucleated vacancy clusters and each nucleated cluster captures a smaller number of defects during the whole irradiation process, inhibiting the growth of vacancy clusters. Above 0.2 dpa, the density of vacancy clusters grows slowly for the existing clusters number is high and act as sinks for movable defects, shown in Fig. 4. Fig. 6 shows the distribution of vacancy cluster sizes at 1.0 dpa. At a higher dose rate, small vacancy clusters less than 2 nm account for a significant proportion and peak at 1.2 nm. With decreasing dose rate, one can find that larger clusters form and distribution of vacancy clusters get broader. These results also demonstrate that the nucleation of vacancy clusters is weakened with decreasing dose rate.

### 3.2 Effect of irradiation particle types

The difference of vacancy cluster characteristics caused by three types of irradiation conditions is also shown in Fig. 3. One can find that the effect of irradiation types is comparatively weaker than that of dose rate. On the whole, the neutron irradiation has the higher density of vacancy clusters than that of the (two types of) ion irradiation in most cases. The divide is up to 50% at 300 °C, $10^{-3}$ dpa/s and decreases with increasing temperature and decreasing dose rate. This is due to the higher proportion of high energy atom (>10 keV) in the PKA spectrum of the neutron irradiation, which introduces more stable vacancy clusters containing more than 5 vacancies and is immobile in our simulation. These results consist with the previous study of PKA energy in Ref. [33] that the number density of vacancy clusters with higher energy cascades is lower than with lower energy cascades. Higher energy cascades of neutron irradiation also result in its smaller average size. However, the irradiation condition of low dose rate and high temperature (400 °C, $10^{-7}$ dpa/s; 500°C, $10^{-5}$ dpa/s) shows the exception. This may come from the limitation of the simulation box size, for the vacancy clusters of the intermedia size have the density lower than the minimal distinguishable density. Additionally, the defect number in the simulation box is too little (~ 10 defects) and make the results less statistical.

In this work, two different ion irradiation types are considered in our simulation: the ion track and the ion spectrum. Ion spectrum condition neglects the spatial structure of the ion irradiations. These two types allow to study the effect of the spatial correlation between the cascades in the ion track. With the ion track simulations, the average size of vacancy clusters is merely 5-10% larger than that of the ion spectrum. This difference is attributed to the more concentrated displacements in the ion track, which promote the aggregation of vacancies and the growth of vacancy clusters.

The number density of vacancy clusters needs more detailed analysis. At low dose range (< 0.2 dpa), the density in ion track condition is higher compared to that in the ion spectrum type. The higher spatial correlation of cascades in the ion track enhances the interaction frequency between mono-vacancies from nearby cascades, and thus promotes the nucleation of vacancy clusters. When the irradiation dose accumulates, the nucleation rate decreases for the majority of newly generated mono-vacancies are absorbed by the existing clusters whose density is far exceed the intrinsic sink. And the number density of ion spectrum increases more rapidly and finally surpass the ion track. With the increase of the temperature or the decrease of the dose rate, the migration events of vacancies increase rapidly and the spatial correlation of nearby cascades becomes less significant. Therefore, the gap of the number density narrows in the types of ion track and ion spectrum. Nevertheless, the impact of irradiation types is less significant than irradiation dose rate.

Fig. 6 shows the size distribution of vacancy clusters under different irradiation particle types. Results above can also be proved by this data: the higher density of vacancy clusters under neutron irradiation mainly comes from the small size clusters around 1nm. And vacancy clusters of larger size forms at the ion track condition because of the spatial correlation of nearby cascades, which results to a lower density and increases its average size. Vacancy clusters large enough to be identified in TEM, i.e., with a diameter larger than approximately 1 nm, and containing more than 50 vacancies per cluster can also be seen in Fig. 6. The void swelling is then calculated by summing the total volume of voids, as shown in Fig. 7. The simulation indicate that the

void swelling is highest in the ion track type, followed by the neutron spectrum, with the ion spectrum exhibiting the lowest swelling under the same irradiation temperature and dose rate. Considering the correlation of cascades in the ion track enhances the swelling mainly by promoting the void growth, with large voids playing the major role in the void swelling.

### 3.3 Effect of irradiation temperature

The evolution of vacancy clusters with irradiation temperature is also shown in Fig. 3. The effect of dose rate and irradiation types remain similar across the three irradiation temperatures. One can also find that as the irradiation temperature increases, the density of vacancy clusters decreases, while their average size increases. These results are in good agreement with the experiment trend in Ref. [59, 60]. This is due to the higher mobility of vacancies and lower stability of small clusters at elevated temperatures, where single vacancies and small mobile clusters tend to coalesce. Additionally, with temperature increasing, the rate of recombination increased (further discussed in the Discussion section), with the increase in the number of vacancy objects that had been absorbed by sinks, simultaneously [61].

Fig. 6 shows the distribution of vacancy cluster sizes at 1.0 dpa, and we can also see the effect of temperature. The largest size of vacancy clusters increases with increasing temperature. At the irradiation condition of low dose rate and high temperature (400 °C, $10^{-7}$ dpa/s; 500 °C, $10^{-5}$ dpa/s), the intermedia size (between 2 nm and 5 nm) clusters vanish, which means its density is lower than the minimal distinguishable density of the simulation box. As a matter of fact, small and intermedia size vacancy clusters become thermally unstable in the higher temperature range and are thus more likely to emit mono-vacancies, which enhances the absorption by large vacancy clusters and expand the size limit of vacancy clusters. Additionally, it also enhances the recombination and absorption by sinks.

## 4. Discussion

Since there is no record of Fe-Cr alloys (or F/M steels) under the experimental conditions of neutron and ion irradiation in a large range of temperatures and dose rates in the literature, and some approximations are used in our OKMC simulation, only the trends of the irradiated microstructure evolution can be compared with experimental observations and draw a conclusion. To obtain better statistics, increasing the dimensions of the simulation system would be useful, especially under the condition of low dose rate and high temperature, but this solution requires much more computational time and is difficult to reach enough damage does, so we just executed every simulation in 3 independent runs.

The density of dislocations for Fe-Cr alloys is around the magnitude of (~$1\times10^{13}$

$m^{-2}$), and slightly growing with increasing Cr content. The effect of dislocation line densities is also studied: three densities, $3.04 \times 10^{14}$ $m^{-2}$, $3.04 \times 10^{13}$ $m^{-2}$, $3.04 \times 10^{12}$ $m^{-2}$ are used in the simulation and the simulations were performed at 300 ℃, 1e-5 dpa/s under varied irradiation conditions. With different dislocation densities, the number density and average size of vacancy clusters under three types of irradiation types are presented in Fig. 8. One can find that the three dislocation densities show little difference on the defect features of vacancy clusters, which means the influence of sink absorption is smaller at a higher dose compared to other irradiation conditions in this general density region of Fe-Cr alloys [19].

To investigate the effect of dose rate at the microscopic level, the interaction of different events in the OKMC simulations was calculated. The average interaction frequency of mono-defects during the whole 1.0 dpa simulation is shown in Fig. 9 (take the neutron spectrum type for example). With the irradiation dose rate increasing by two orders of magnitude, the interaction frequency of mono-defects increases near two orders of magnitude. This is of course due to the larger number density of defects introduced per unit time and volume and the nucleation is enhanced at a higher dose rate. The movable vacancy defects are more likely to interact with nearby defects and form small, stable clusters with a higher number density. On the contrary, at a lower dose rate, with a lower interaction frequency, the movable defects can migrate with a larger mean free path and form larger vacancy clusters. This partly explains the dose rate effect that high dose rate enhances the nucleation of vacancy clusters.

The simulation results in Fig. 10 show the total number of events that vacancy clusters interact with (a-c) mono-vacancies and vacancy clusters, and (d-f) mono-interstitials and SIA clusters during the whole simulation of 1.0 dpa. These results also support the dose rate effect. The former interactions promote the growth of vacancy clusters while the latter inhibit the growth. For the same type of irradiation, with the dose rate decreasing and the temperature increasing, the number of reaction events with vacancy clusters increase, while the interactions with SIA clusters decrease simultaneously. This indicates that more vacancies can be captured by one vacancy cluster and the vacancy cluster coalescence is enhanced, which results in the larger average size of vacancy clusters. According to the statistical results of other events, the effects of thermal dissociation and sink absorption are much smaller than the events described above.

The impact of irradiation particle types is also shown in Fig. 10. At the same temperature and dose rate, the number of reaction events between vacancy clusters and vacancy (clusters), as well as reaction events between vacancy clusters and SIA (clusters) follows the same patterns: the neutron spectrum type results in the highest number of both events, followed by the ion track, with the ion spectrum type exhibiting the smallest number of events. Consequently, with the changing of irradiation particle types, the effect of irradiation types on vacancy cluster density and average size is diminished, as the two types of events offset each other. Overall, the impact of irradiation particle types is less significant than irradiation temperature and dose rate.

Fig. 11 shows the absorption of mono-defects by dislocations. The absorption number increasing with increasing temperature for the high mobilities of defects at high temperatures. At the same temperature, the higher absorption number also means more migration events for vacancies and SIAs. This also verifies that a lower dose rate results to a longer migration path of defects before interacting with other defects, for its high absorption of mono-defects by intrinsic sinks. As for the three different irradiation types, the defects of ion track type have a high probability absorbed by dislocations, consistent with its large cluster size and low number density of vacancy clusters.

**5. Conclusions**

In this work, we proposed object kinetic Monte Carlo simulations to investigate the evolution of the microstructure in Fe-9%Cr alloys. Cascade morphologies of different types of fission-neutron and heavy-ion irradiations were constructed to investigate the influence of PKA energy spectra and spatial distribution of displacement cascades. The effects of dose rate, temperature and irradiation type on vacancy cluster evolution were studied. Our simulation shows that the density of vacancy clusters increases with increasing dose rate, while the average size decreases. High dose rate greatly promotes the interaction frequency between small defects by orders of magnitude and thus enhance the nucleation of vacancy clusters. Spontaneously, the absorption of small vacancies by one vacancy cluster is also inhibited. This dose rate effect caused by microscopic interaction of small mobile defects explains the major difference of microstructure between fission neutron and heavy ion irradiation. The density of vacancy clusters decreases with increasing temperature, while the average size increases simultaneously. The distribution of vacancy cluster sizes is also statistically analyzed and the larger clusters form and distribution of vacancy clusters get broader with increasing temperature. This can be attributed to the high mobilities of irradiation defects at high temperatures, and events for vacancy clusters interacting with vacancies increase, while the interactions with SIAs decrease simultaneously, which results to a higher diameter of clusters.

The irradiation particle types also influence the microstructure of vacancy clusters primarily due to differences in the PKA energy spectra and spatial distribution of displacement cascades. Consequently, the impact of irradiation particle types is less significant than irradiation dose rate and temperature. Under the same irradiation dose rate and temperature, neutron irradiation results in the largest number density and smallest average size among the three irradiation types.

In this work, OKMC simulations that provide spatial information of primary defects were performed to investigate the effect of dose rate, irradiation temperature, PKA energy spectra and spatial distribution of displacements cascades on the microstructure of vacancy cluster. These simulations explain the effect of dose rate and temperature, and give deeper knowledge of the effect of spatial correlation of cascades in ion irradiations. Moreover, this work provides new insights into the differences between fission neutron and heavy ion irradiation irradiations on irradiation microstructure evolution in F/M steels.

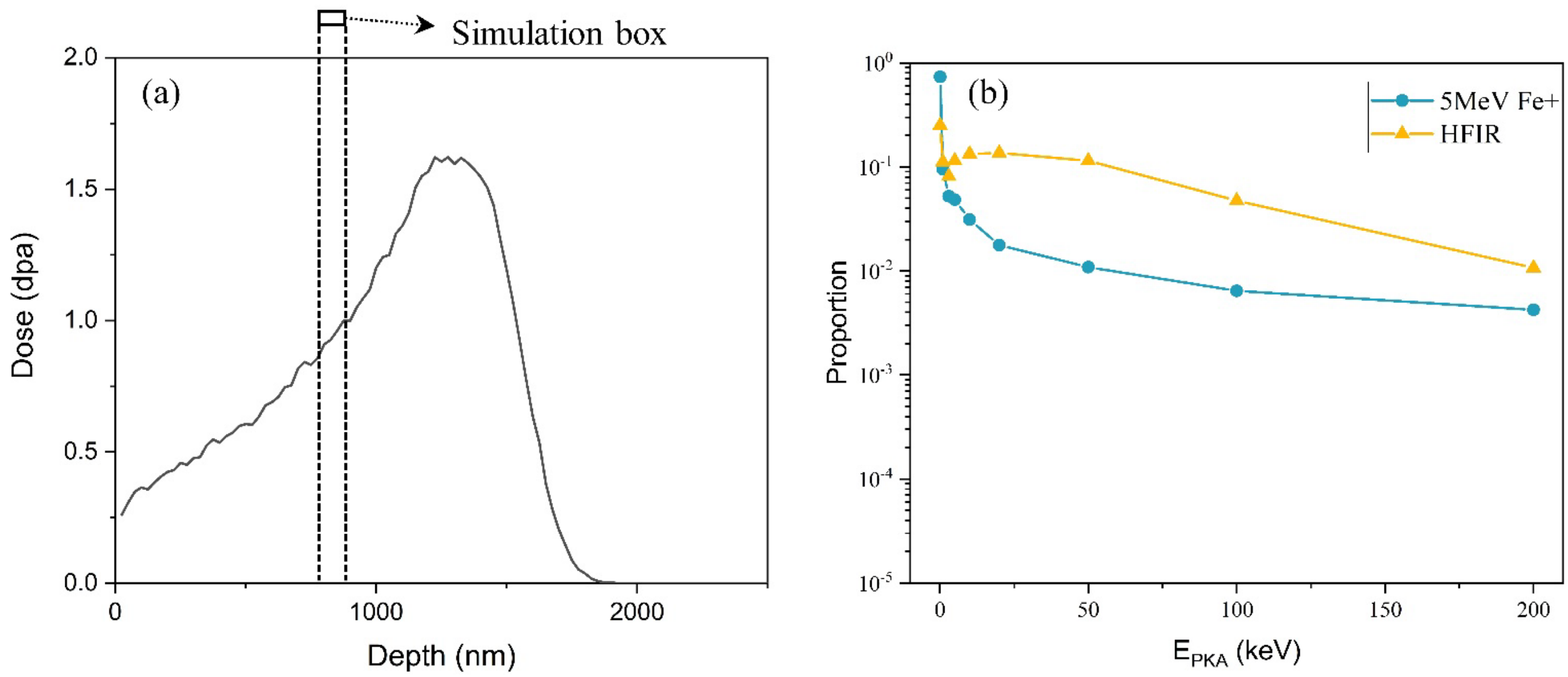

**Fig. 1.** (a) Analysis area of the 5 MeV ion irradiation and the simulation box size. (b) PKA energy spectra of the HFIR reactor and 5 MeV Fe irradiation. Dots show for what values debris were obtained with MD and inserted in the model.

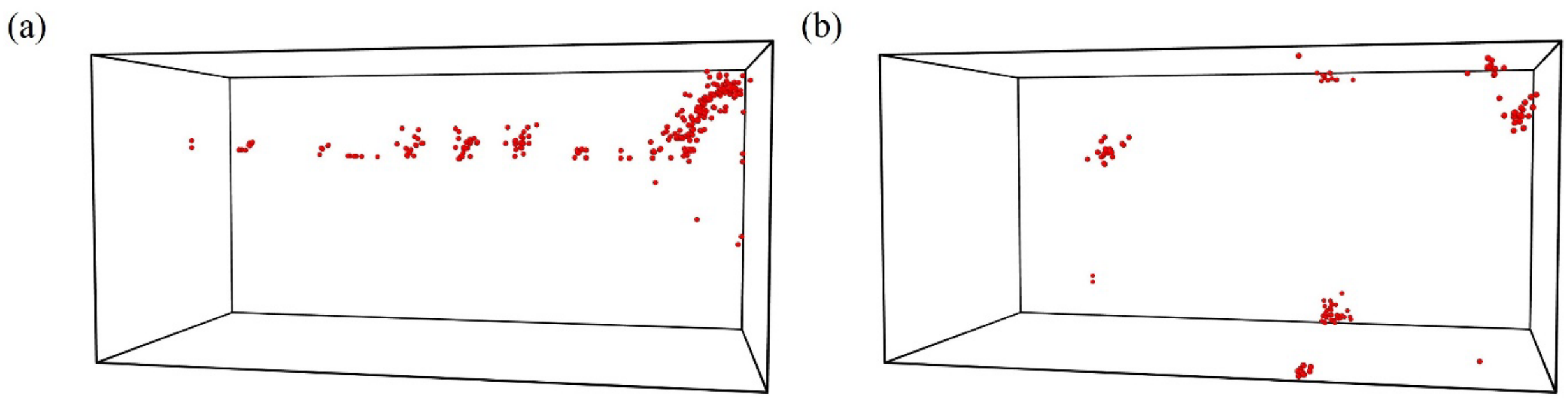

**Fig. 2.** The displacement cascades morphology of different irradiation particles: (a) ion irradiation (track of cascades); (b) neutron irradiation (randomly distributed).

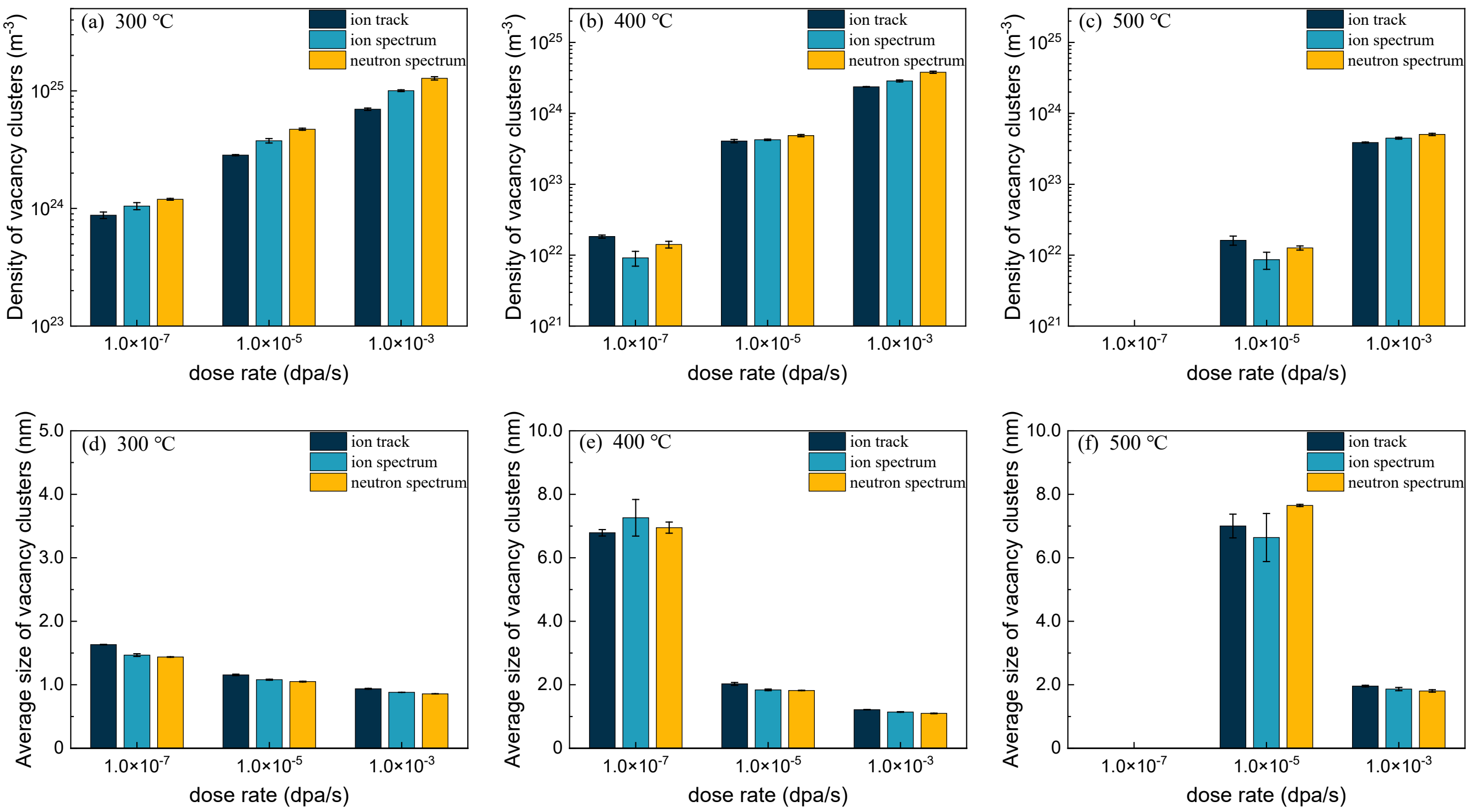

**Fig. 3.** The number density of vacancy clusters for three types of irradiation particles versus dose rate at 1.0 dpa in irradiation temperature (a) 300 ℃; (b) 400 ℃; (c) 500 ℃. The average size of vacancy clusters in (d) 300 ℃; (e) 400 ℃; (f) 500 ℃.

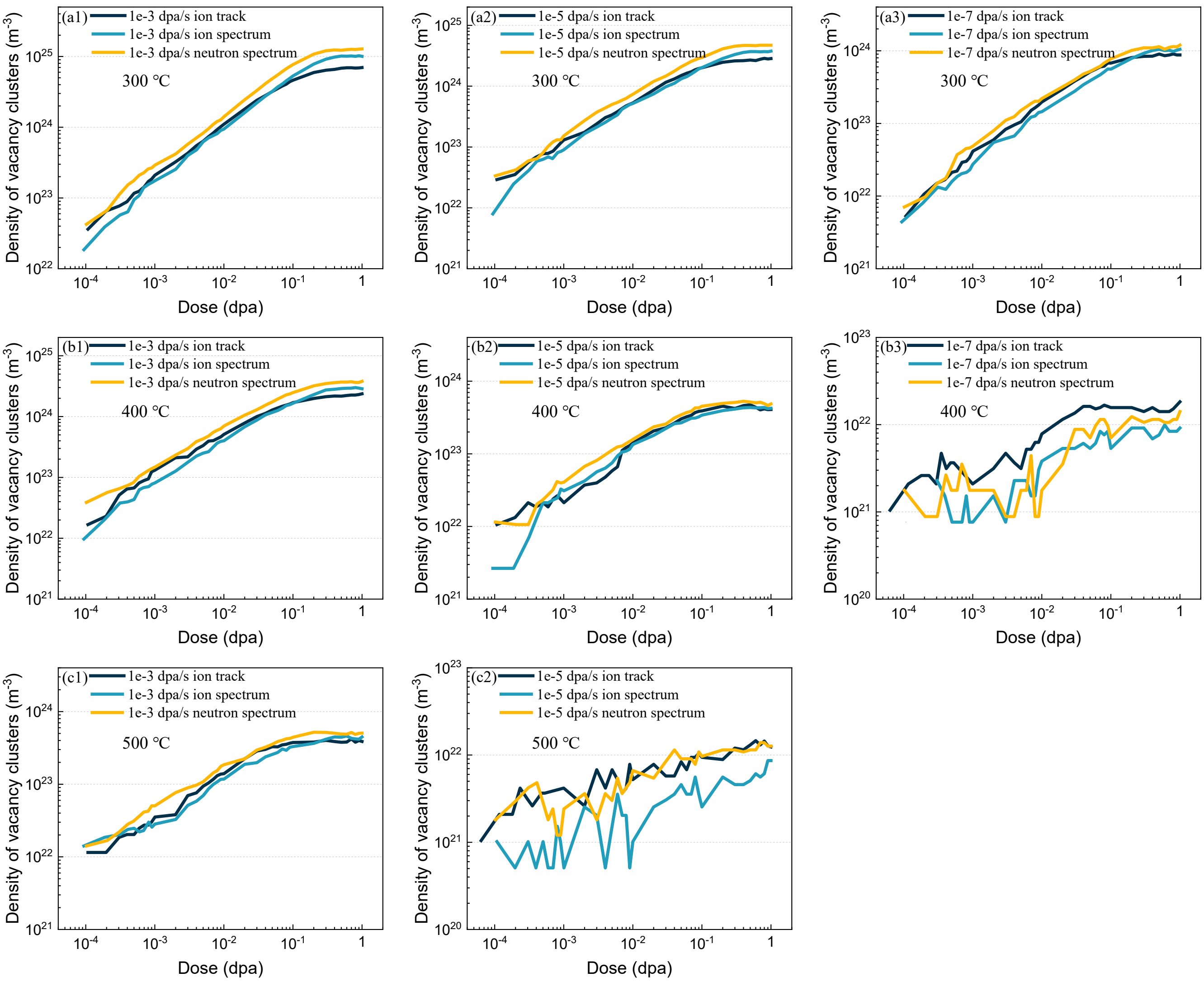

**Fig. 4.** The number density of vacancy clusters for three types of irradiation particles under three different dose rates and irradiation temperatures of (a1-a3) 300 °C; (b1-b3) 400 °C; (c1-c3) 500 °C.

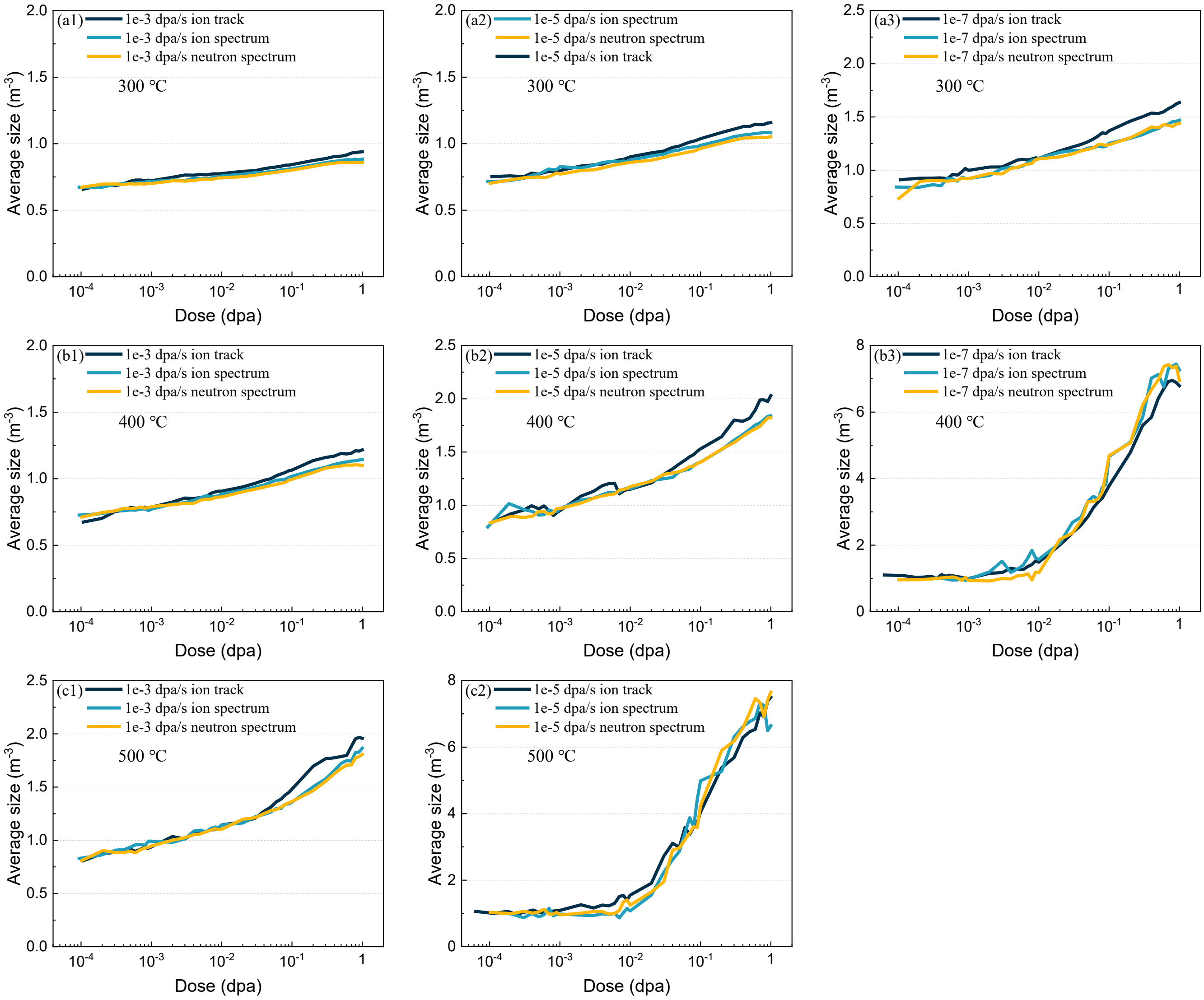

**Fig. 5.** The average size of vacancy clusters for three types of irradiation particles under three different dose rates and irradiation temperatures of (a1-a3) 300 °C; (b1-b3) 400 °C; (c1-c3) 500 °C.

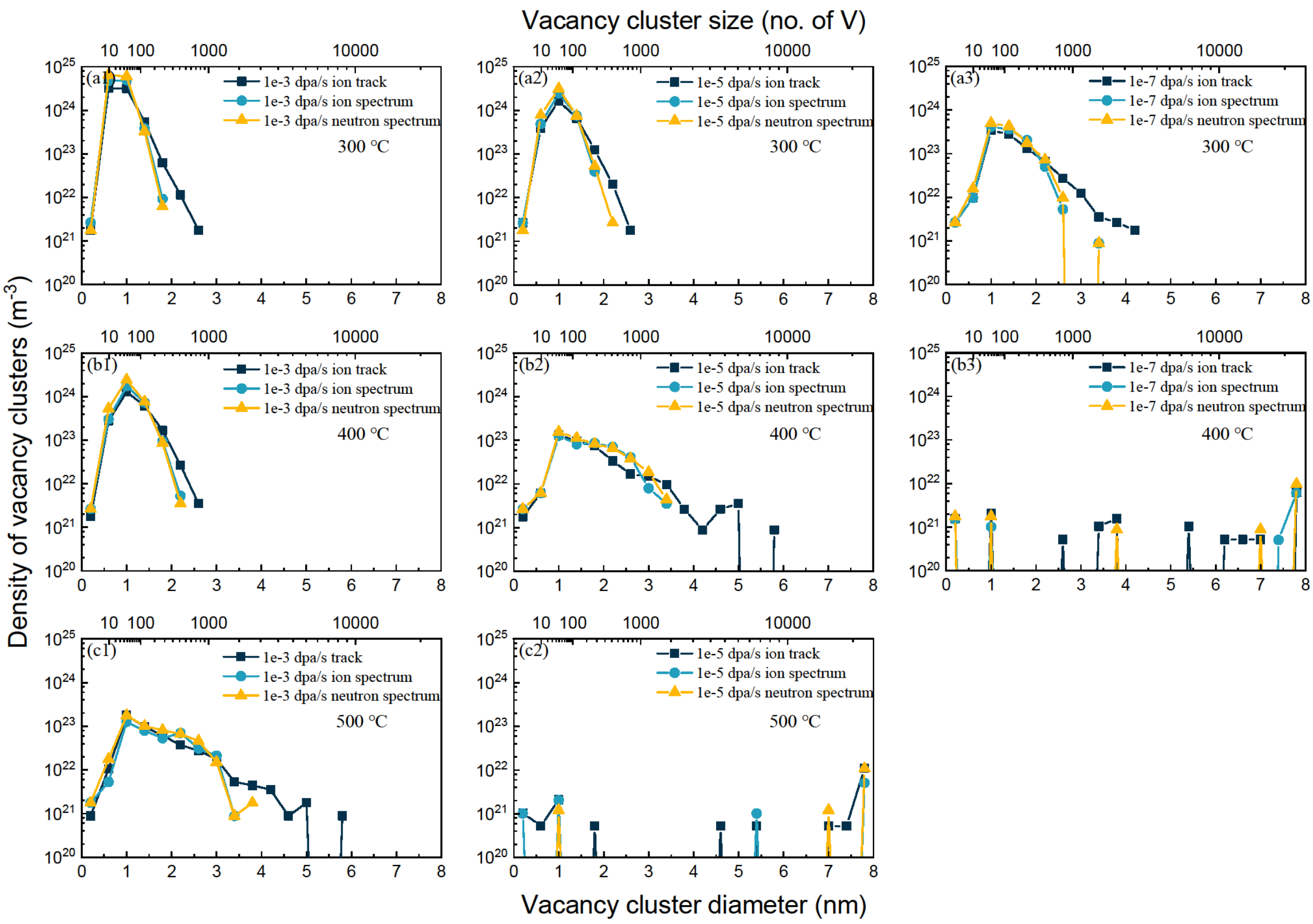

**Fig. 6.** Distribution of sizes for the vacancy clusters in the OKMC simulation box, at three irradiation particles, three different dose rates and irradiation temperatures of (a1-a3) 300 °C; (b1-b3) 400 °C; (c1-c3) 500 °C.

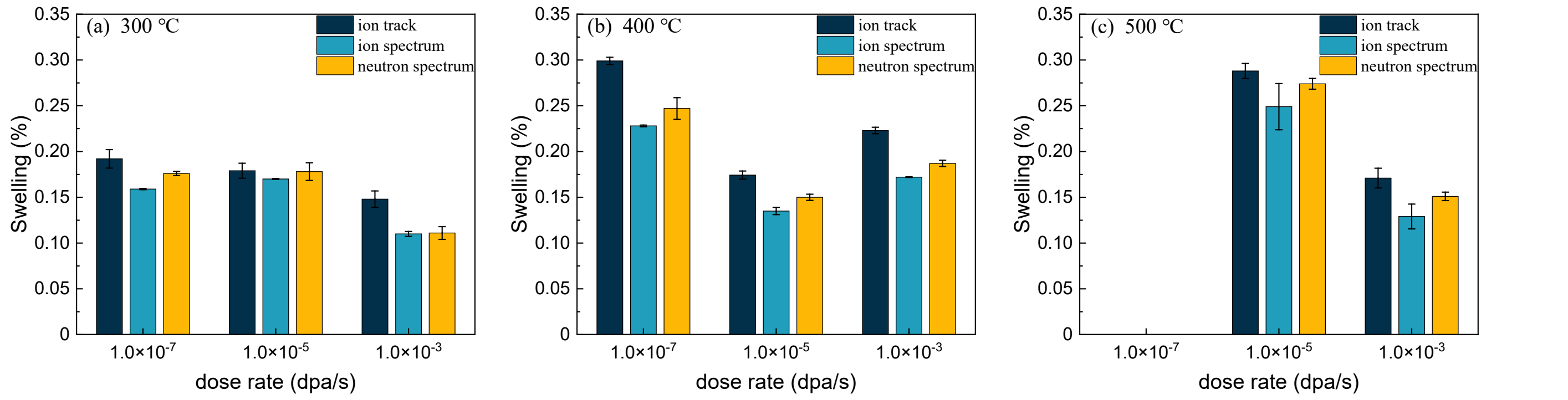

**Fig. 7.** The void swelling for three types of irradiation particles versus dose rate at 1.0 dpa and irradiation temperature (a) 300 °C; (b) 400 °C; (c) 500 °C.

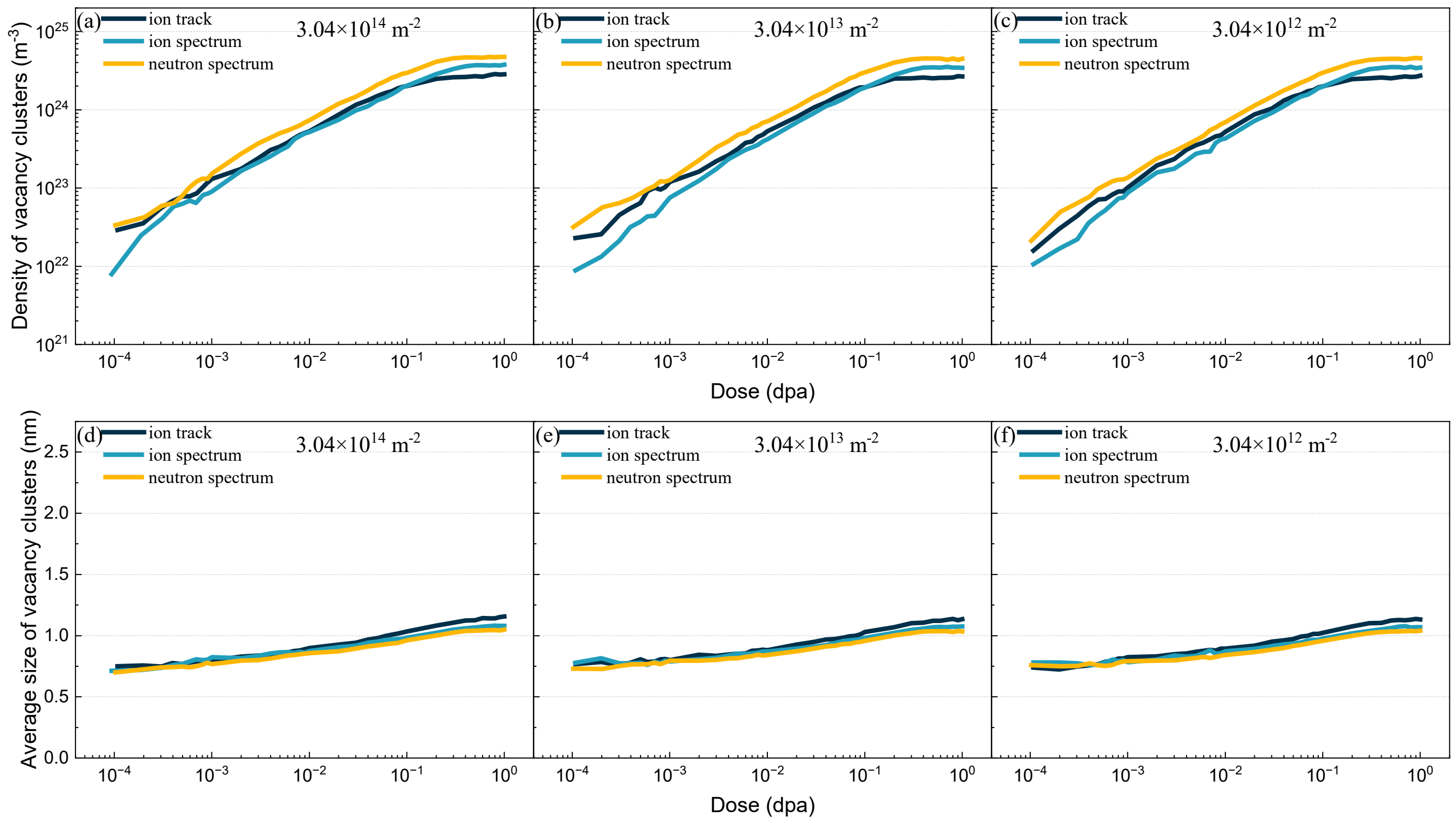

**Fig. 8.** The number density of vacancy clusters under dislocation densities of (a) $3.04 \times 10^{14}$ m$^{-2}$, (b) $3.04 \times 10^{13}$ m$^{-2}$, (c) $3.04 \times 10^{12}$ m$^{-2}$. The average size of vacancy clusters under dislocation densities of (d) $3.04 \times 10^{14}$ m$^{-2}$, (e) $3.04 \times 10^{13}$ m$^{-2}$, (f) $3.04 \times 10^{12}$ m$^{-2}$, at irradiation temperature 300 °C, 1e-5 dpa/s.

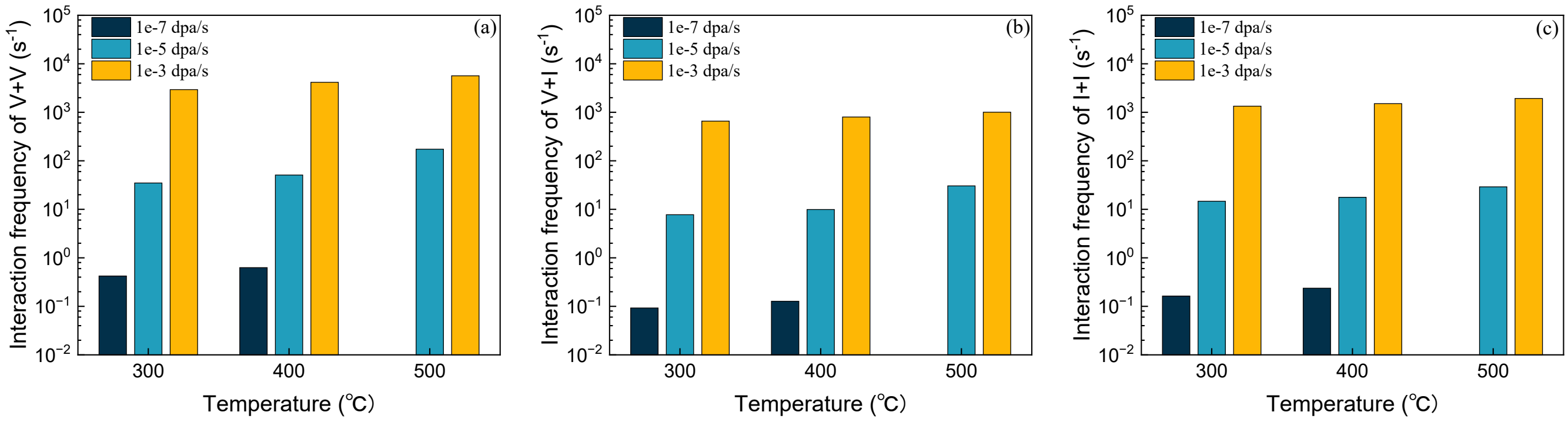

**Fig. 9.** The interaction frequency of mono-defects: (a) V+V; (b) V+I; (c) I+I for neutron irradiation type.

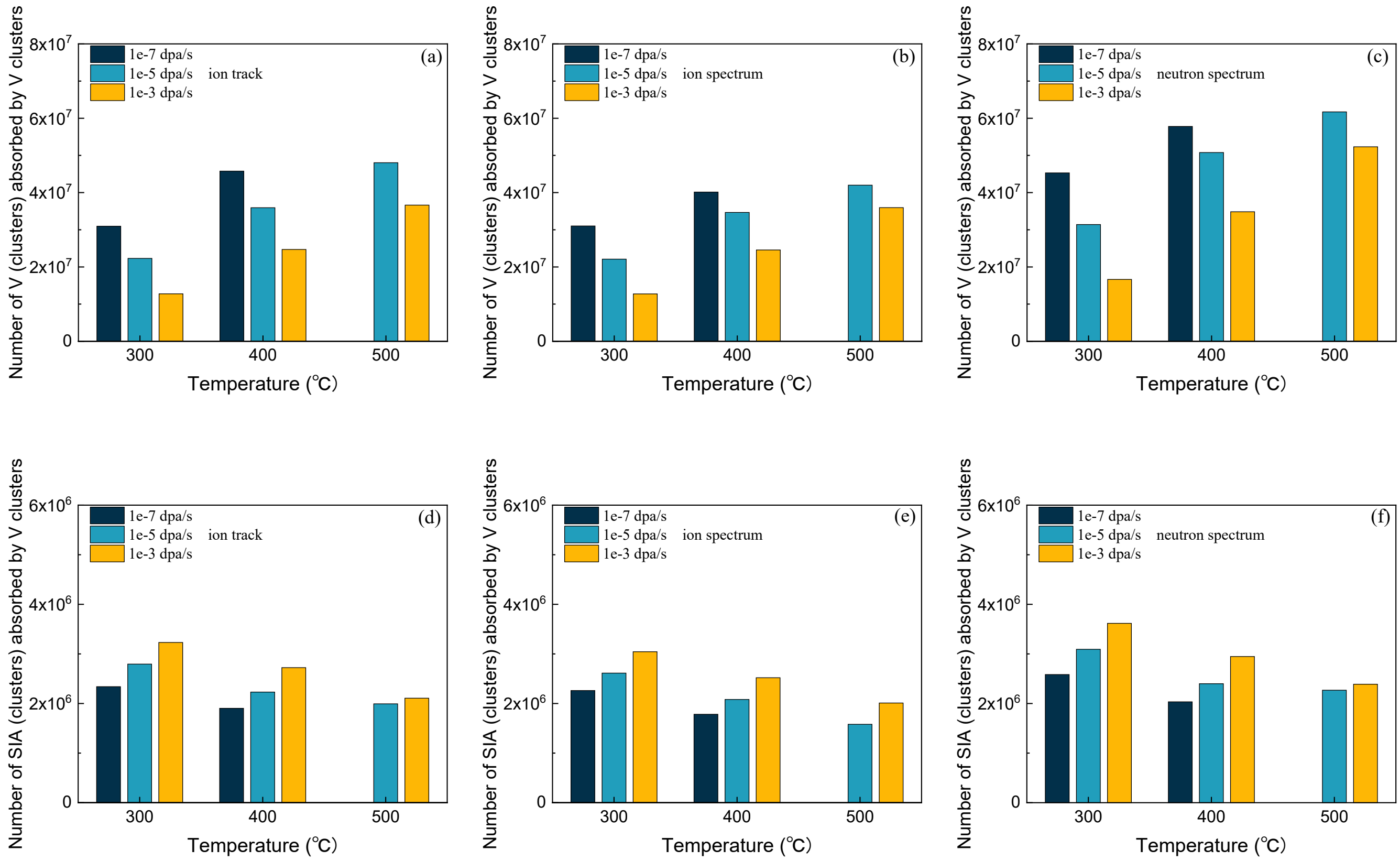

**Fig. 10.** The number of events that (a-c) mono-vacancies and vacancy clusters and (d-f) mono-interstitials and SIA clusters interact with vacancy clusters under different irradiation conditions at 1.0 dpa, respectively.

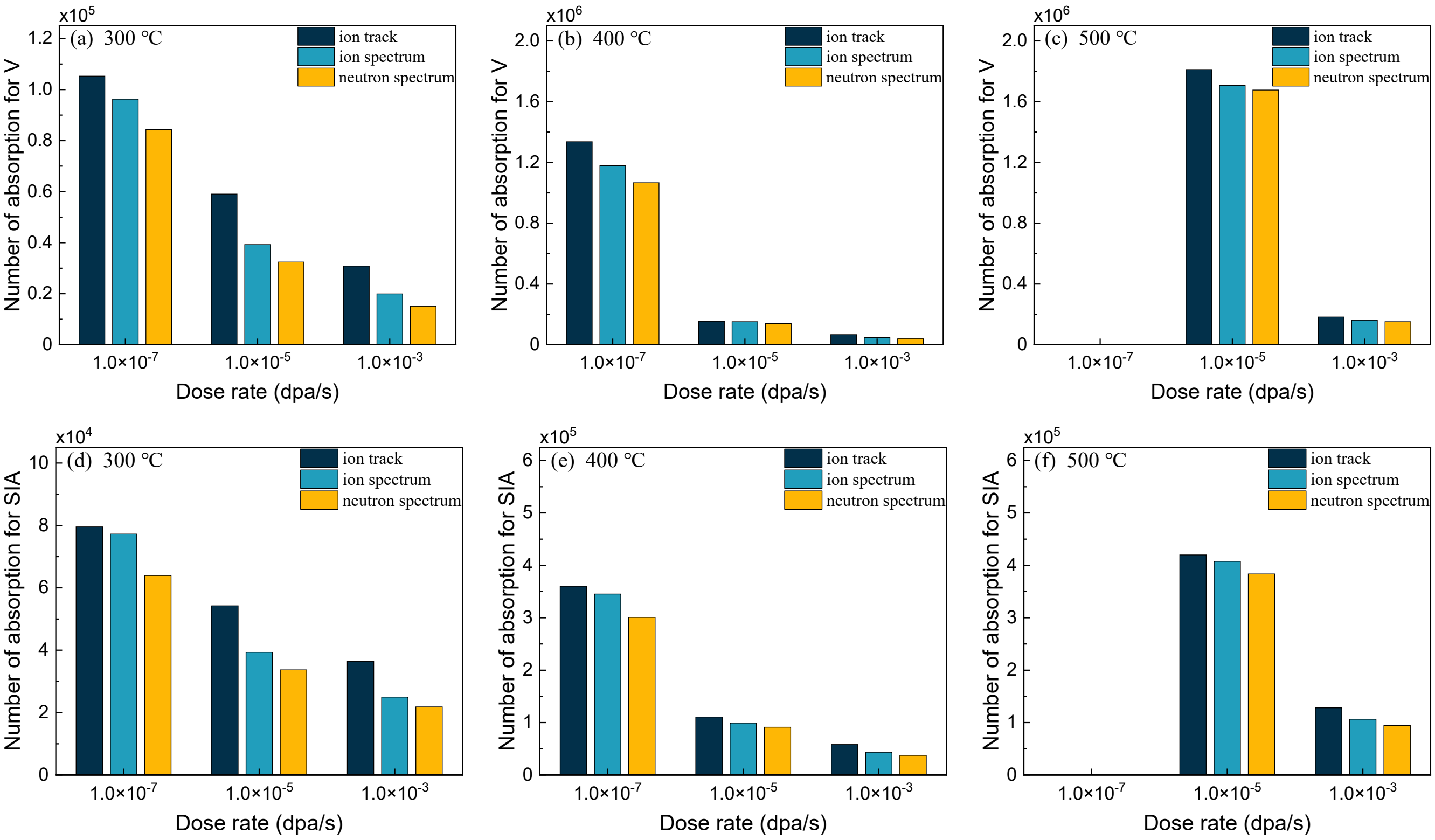

**Fig. 11.** The number of (a-c) mono-vacancies (d-f) mono-SIAs absorbed by dislocations under different irradiation conditions at 1.0 dpa, respectively.

[1] S.J. Zinkle, L.L. Snead, Designing Radiation Resistance in Materials for Fusion Energy*, 44(Volume 44, 2014) (2014) 241-267.
[2] N.H. Packan, K. Farrell, J.O. Stiegler, Correlation of neutron and heavy-ion damage: I. The influence of dose rate and injected helium on swelling in pure nickel, Journal of Nuclear Materials 78(1) (1978) 143-155.
[3] L.K. Mansur, Correlation of neutron and heavy-ion damage: II. The predicted temperature shift if swelling with changes in radiation dose rate, Journal of Nuclear Materials 78(1) (1978) 156-160.
[4] Accelerator Simulation and Theoretical Modelling of Radiation Effects in Structural Materials, INTERNATIONAL ATOMIC ENERGY AGENCY, Vienna, 2018.
[5] B.J. Heidrich, S.M. Pimblott, G.S. Was, S.J.J.N.I. Zinkle, M.i.P.R.S.B.B.I.w. Materials, Atoms, Roadmap for the application of ion beam technologies to the challenges of nuclear energy technologies, (2019).
[6] J.G. Gigax, E. Aydogan, T. Chen, D. Chen, L. Shao, Y. Wu, W.Y. Lo, Y. Yang, F.A. Garner, The influence of ion beam rastering on the swelling of self-ion irradiated pure iron at 450 °C, Journal of Nuclear Materials 465 (2015) 343-348.
[7] Z. Jiao, J. Michalicka, G.S. Was, Self-ion emulation of high dose neutron irradiated microstructure in stainless steels, Journal of Nuclear Materials 501 (2018) 312-318.
[8] S.J. Zinkle, L.L. Snead, Opportunities and limitations for ion beams in radiation effects studies: Bridging critical gaps between charged particle and neutron irradiations, Scripta Materialia 143 (2018) 154-160.
[9] G.S. Was, Z. Jiao, E. Getto, K. Sun, A.M. Monterrosa, S.A. Maloy, O. Anderoglu, B.H. Sencer, M. Hackett, Emulation of reactor irradiation damage using ion beams, Scripta Materialia 88 (2014) 33-36.
[10] G.S. Was, Challenges to the use of ion irradiation for emulating reactor irradiation, Journal of Materials Research 30(9) (2015) 1158-1182.
[11] G.S. Was, T.R. Allen, RADIATION DAMAGE FROM DIFFERENT PARTICLE TYPES, in: K.E. Sickafus, E.A. Kotomin, B.P. Uberuaga (Eds.) Radiation Effects in Solids, Springer Netherlands, Dordrecht, 2007, pp. 65-98.
[12] A. Ballesteros, R. Ahlstrand, C. Bruynooghe, A. Chernobaeva, Y. Kevorkyan, D. Erak, D. Zurko, Irradiation temperature, flux and spectrum effects, Progress in Nuclear Energy 53(6) (2011) 756-759.
[13] R.E. Stoller, G.R. Odette, B.D. Wirth, Primary damage formation in bcc iron, Journal of Nuclear Materials 251 (1997) 49-60.
[14] N. Castin, G. Bonny, A. Bakaev, C.J. Ortiz, A.E. Sand, D. Terentyev, Object kinetic Monte Carlo model for neutron and ion irradiation in tungsten: Impact of transmutation and carbon impurities, Journal of Nuclear Materials 500 (2018) 15-25.
[15] A.D. Brailsford, R. Bullough, The rate theory of swelling due to void growth in irradiated metals, Journal of Nuclear Materials 44(2) (1972) 121-135.
[16] L.K. Mansur, Void Swelling in Metals and Alloys Under Irradiation: An Assessment of the Theory, Nuclear Technology 40(1) (1978) 5-34.
[17] G.S. Was, J.T. Busby, T. Allen, E.A. Kenik, A. Jensson, S.M. Bruemmer, J. Gan, A.D. Edwards, P.M. Scott, P.L. Andreson, Emulation of neutron irradiation effects with protons: validation of principle, Journal of Nuclear Materials 300(2) (2002) 198-216.

[18] M. Chiapetto, C.S. Becquart, L. Malerba, Simulation of nanostructural evolution under irradiation in Fe-9%CrC alloys: An object kinetic Monte Carlo study of the effect of temperature and dose-rate, Nuclear Materials and Energy 9 (2016) 565-570.
[19] J. Li, C. Zhang, Y. Yang, T. Wang, I. Martin-Bragado, Irradiation dose-rate effect in Fe-C system: An Object Kinetic Monte Carlo simulation, Journal of Nuclear Materials 561 (2022).
[20] S. Taller, G. VanCoevering, B.D. Wirth, G.S. Was, Predicting structural material degradation in advanced nuclear reactors with ion irradiation, Sci Rep 11(1) (2021) 2949.
[21] L.N. Clowers, G.S. Was, The effect of hydrogen co-injection on the microstructure of triple ion irradiated F82H, Journal of Nuclear Materials 601 (2024) 155331.
[22] L.N. Clowers, Z. Jiao, G.S. Was, Synergies between H, He and radiation damage in dual and triple ion irradiation of candidate fusion blanket materials, Journal of Nuclear Materials 565 (2022) 153722.
[23] Y. Li, A. French, Z. Hu, A. Gabriel, L.R. Hawkins, F.A. Garner, L. Shao, A quantitative method to determine the region not influenced by injected interstitial and surface effects during void swelling in ion-irradiated metals, Journal of Nuclear Materials 573 (2023) 154140.
[24] Z. Hu, L. Shao, Effects of Carbon on Void Nucleation in Self-Ion–Irradiated Pure Iron, Nuclear Science and Engineering 198(1) (2024) 145-157.
[25] S.J.J.C.N.M. Zinkle, Radiation-Induced Effects on Microstructure, (2020).
[26] R.F. Mattas, F.A. Garner, M.L. Grossbeck, P.J. Maziasz, G.R. Odette, R.E. Stoller, The impact of swelling on fusion reactor first wall lifetime, Journal of Nuclear Materials 122(1) (1984) 230-235.
[27] H. Tanigawa, K. Shiba, A. Möslang, R.E. Stoller, R. Lindau, M.A. Sokolov, G.R. Odette, R.J. Kurtz, S. Jitsukawa, Status and key issues of reduced activation ferritic/martensitic steels as the structural material for a DEMO blanket, Journal of Nuclear Materials 417(1) (2011) 9-15.
[28] D.S. Gelles, Microstructural examination of commercial ferritic alloys at 200 dpa, Journal of Nuclear Materials 233-237 (1996) 293-298.
[29] R.L. Klueh, D.R. Harries, High-Chromium Ferritic and Martensitic Steels for Nuclear Applications, 2001.
[30] L. Malerba, A. Caro, J. Wallenius, Multiscale modelling of radiation damage and phase transformations: The challenge of FeCr alloys, Journal of Nuclear Materials 382(2) (2008) 112-125.
[31] M. Chiapetto, L. Malerba, C.S. Becquart, Effect of Cr content on the nanostructural evolution of irradiated ferritic/martensitic alloys: An object kinetic Monte Carlo model, Journal of Nuclear Materials 465 (2015) 326-336.
[32] S.J. Zinkle, R.E. Stoller, Quantifying defect production in solids at finite temperatures: Thermally-activated correlated defect recombination corrections to DPA (CRC-DPA), Journal of Nuclear Materials 577 (2023).
[33] A. Souidi, M. Hou, C.S. Becquart, L. Malerba, C. Domain, R.E. Stoller, On the correlation between primary damage and long-term nanostructural evolution in iron under irradiation, Journal of Nuclear Materials 419(1) (2011) 122-133.
[34] C.-C. Fu, J.D. Torre, F. Willaime, J.-L. Bocquet, A. Barbu, Multiscale modelling of defect kinetics in irradiated iron, Nature Materials 4(1) (2004) 68-74.

[35] C. Domain, C.S. Becquart, L. Malerba, Simulation of radiation damage in Fe alloys: an object kinetic Monte Carlo approach, Journal of Nuclear Materials 335(1) (2004) 121-145.
[36] V. Jansson, M. Chiapetto, L. Malerba, The nanostructure evolution in Fe–C systems under irradiation at 560K, Journal of Nuclear Materials 442(1) (2013) 341-349.
[37] G.S. Was, The Damage Cascade, in: G.S. Was (Ed.), Fundamentals of Radiation Materials Science: Metals and Alloys, Springer New York, New York, NY, 2017, pp. 131-165.
[38] I. Martin-Bragado, A. Rivera, G. Valles, J.L. Gomez-Selles, M.J. Caturla, MMonCa: An Object Kinetic Monte Carlo simulator for damage irradiation evolution and defect diffusion, (2013).
[39] W.M. Young, E.W. Elcock, Monte Carlo studies of vacancy migration in binary ordered alloys: I, Proceedings of the Physical Society 89(3) (1966) 735.
[40] A.B. Bortz, M.H. Kalos, J.L. Lebowitz, A new algorithm for Monte Carlo simulation of Ising spin systems, Journal of Computational Physics 17(1) (1975) 10-18.
[41] L. Malerba, C.S. Becquart, C. Domain, Object kinetic Monte Carlo study of sink strengths, Journal of Nuclear Materials 360(2) (2007) 159-169.
[42] N. Soneda, S. Ishino, A. Takahashi, K. Dohi, Modeling the microstructural evolution in bcc-Fe during irradiation using kinetic Monte Carlo computer simulation, Journal of Nuclear Materials 323(2) (2003) 169-180.
[43] J.P. Balbuena, M.J. Aliaga, I. Dopico, M. Hernández-Mayoral, L. Malerba, I. Martin-Bragado, M.J. Caturla, Insights from atomistic models on loop nucleation and growth in α-Fe thin films under Fe+ 100 keV irradiation, Journal of Nuclear Materials 521 (2019) 71-80.
[44] D. Terentyev, I. Martin-Bragado, Evolution of dislocation loops in iron under irradiation: The impact of carbon, Scripta Materialia 97 (2015) 5-8.
[45] J. Marian, B.D. Wirth, J.M. Perlado, Mechanism of Formation and Growth of $〈100〉$ Interstitial Loops in Ferritic Materials, Physical Review Letters 88(25) (2002) 255507.
[46] C. Björkas, K. Nordlund, M.J. Caturla, Influence of the picosecond defect distribution on damage accumulation in irradiated α-Fe, Physical Review B 85(2) (2012).
[47] F. Luo, B. Zhang, Z. Gao, J. Huang, H.-B. Zhou, G.-H. Lu, F. Gao, Y. Wang, C. Wang, Quantitative method to predict the energetics of helium-nanocavities interactions in metal systems based on electrophobic interaction, Journal of Materiomics 10(3) (2024) 725-737.
[48] R. Alexander, M.C. Marinica, L. Proville, F. Willaime, K. Arakawa, M.R. Gilbert, S.L. Dudarev, Ab initioscaling laws for the formation energy of nanosized interstitial defect clusters in iron, tungsten, and vanadium, Physical Review B 94(2) (2016).
[49] M. Jaraiz, G.H. Gilmer, D.M. Stock, T. Diaz de la Rubia, Defects from implantation in silicon by linked Marlow-molecular dynamics calculations, Nuclear Instruments and Methods in Physics Research Section B: Beam Interactions with Materials and Atoms 102(1) (1995) 180-182.
[50] C.J. Ortiz, A combined BCA-MD method with adaptive volume to simulate high-energy atomic-collision cascades in solids under irradiation, Computational Materials Science 154 (2018) 325-334.
[51] A. De Backer, A. Sand, C.J. Ortiz, C. Domain, P. Olsson, E. Berthod, C.S. Becquart, Primary damage in tungsten using the binary collision approximation, molecular dynamic simulations and the density functional theory, Physica Scripta 2016(T167) (2016) 014018.
[52] J. Hou, X. Kong, W. Hu, H. Deng, D. Nguyen-Manh, J. Song, Deuterium trapping and

desorption by vacancy clusters in irradiated Mo from object kinetic Monte Carlo simulations, Acta Materialia 274 (2024) 120014.
[53] J.F. Ziegler, M.D. Ziegler, J.P. Biersack, SRIM – The stopping and range of ions in matter (2010), Nuclear Instruments and Methods in Physics Research Section B: Beam Interactions with Materials and Atoms 268(11) (2010) 1818-1823.
[54] FISPACT-II Wiki Home Page, https://fispact.ukaea.uk/wiki/Main_Page#Resources_for_users, (n.d.).
[55] S. Plimpton, Fast Parallel Algorithms for Short-Range Molecular Dynamics, Journal of Computational Physics 117(1) (1995) 1-19.
[56] L. Malerba, M.C. Marinica, N. Anento, C. Björkas, H. Nguyen, C. Domain, F. Djurabekova, P. Olsson, K. Nordlund, A. Serra, D. Terentyev, F. Willaime, C.S. Becquart, Comparison of empirical interatomic potentials for iron applied to radiation damage studies, Journal of Nuclear Materials 406(1) (2010) 19-38.
[57] M.C. Marinica, F. Willaime, J.P. Crocombette, Irradiation-Induced Formation of Nanocrystallites with $C15$ Laves Phase Structure in bcc Iron, Physical Review Letters 108(2) (2012) 025501.
[58] M.J. Norgett, M.T. Robinson, I.M. Torrens, A proposed method of calculating displacement dose rates, Nuclear Engineering and Design 33(1) (1975) 50-54.
[59] L.L. Horton, J. Bentley, K. Farrell, A TEM study of neutron-irradiated iron, Journal of Nuclear Materials 108-109 (1982) 222-233.
[60] T. Muroga, H. Watanabe, N. Yoshida, Correlation of fast neutron, fusion neutron and electron irradiations based on the dislocation loop density, Journal of Nuclear Materials 174(2) (1990) 282-288.
[61] H.-Z. Ma, Y.-H. Li, Y.-Z. Niu, D. Terentyev, Z. Yang, H.-B. Zhou, G.-H. Lu, Initial microstructure and temperature dependence of irradiation defects evolution in tungsten, Journal of Nuclear Materials 591 (2024) 154932.